\documentclass[aps,prb,twocolumn,groupedaddress,floatfix,showpacs]{revtex4}
\usepackage{graphicx}
\usepackage{amssymb}
\usepackage{amsmath}

\ifx\pdftexversion\undefined
\usepackage[dvips]{hyperref}
\else
\usepackage{hyperref}
\fi
\hypersetup{
  colorlinks = true, linkcolor = blue
}
\begin{document}

\title{Dynamical properties of ultracold bosons in an optical lattice}

\author{S.D.\ Huber$^{1}$}
\author{E. Altman$^{2}$}
\author{H.P.\ B\"uchler$^{3,4}$}
\author{G. Blatter$^{1}$}

\affiliation{$^{1}$Theoretische Physik, ETH Zurich, CH-8093
Zurich, Switzerland}
\affiliation{$^{2}$ Department of Condensed Matter Physics, The Weizmann
Institute of Science Rehovot, 76100, Israel}
\affiliation{$^{3}$Institute for Quantum Optics and Quantum Information of the
Austrian Academy of Science, 6020 Innsbruck, Austria }
\affiliation{$^{4}$Institut f\"ur theoretische Physik, Universit\"at
Innsbruck, 6020 Innsbruck, Austria}

\date{\today}

\begin{abstract} 
  
We study the excitation spectrum of strongly correlated lattice bosons for the
Mott-insulating phase and for the superfluid phase close to localization.
Within a Schwinger-boson mean-field approach we find two gapped modes in the
Mott insulator and the combination of a sound mode (Goldstone) and a gapped
(Higgs) mode in the superfluid. To make our findings comparable with
experimental results, we calculate the dynamic structure factor as well as the
linear response to the optical lattice modulation introduced by St\"oferle
{\it et al.} [Phys.\ Rev.\ Lett. {\bf 92}, 130403 (2004)]. We find that the
puzzling finite frequency absorption observed in the superfluid phase could be
explained via the excitation of the gapped (Higgs) mode. We check the
consistency of our results with an adapted $f$-sum-rule and  propose an
extension of the experimental technique by  St\"oferle {\it et al.} to further
verify our findings.

\end{abstract}


\maketitle

%%%%%%%%%%%%%%%%%%%%%%%%%%%%%%%%%%%%%%%%%
\section{Introduction}

The prime example of a strongly correlated Bose system is $^{4}$He --- it
exhibits normal, superfluid, crystalline, and possibly even supersolid
phases\cite{Andreev69,Chester70,Leggett70} and is attracting interest to this
day.\cite{Kim04} {\it Dilute} cold bosonic atoms reside generically in the
weakly interacting limit but their tunability through quantum optical
techniques allows for the realization of strongly correlated states.  One way
to achieve the strongly interacting limit is to reduce the kinetic energy by
the application of an optical lattice, thereby effectively enhancing the
effects of interactions. The broken translational symmetry then leads to new
effects in the superfluid phase not present in $^{4}$He. The present paper is
devoted to a study of the dynamical properties of lattice bosons within this
strongly correlated regime.
 
\begin{figure}[b] 
\begin{center} 
\includegraphics{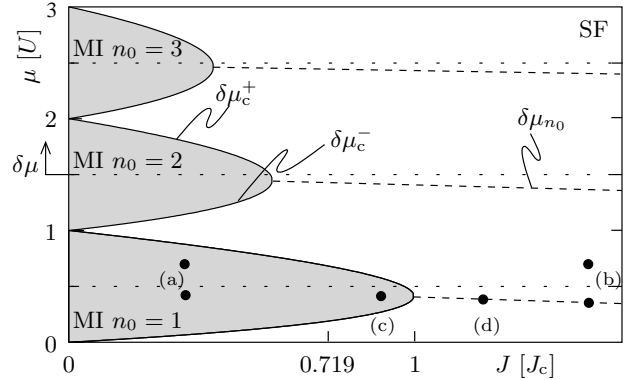} 
\end{center}
\caption{ \label{fig:lobes}
The mean-field phase diagram of the Bose-Hubbard model involves disconnected
incompressible Mott-insulating phases (grey shaded areas at small hopping $J$)
where the density is pinned at integer values $n_{0}$, and a connected
superfluid phase in between and at larger values of $J$. The lines
$\delta\mu_{\rm c}^{\pm}$ mark the second order quantum phase transition
separating these phases.  The commensurate filling in the Mott lobes is
exported into the superfluid along the lines $\delta\mu_{n_{0}}$ (bent
downward) where particle-hole symmetry is preserved.  In our analysis we use a
truncation scheme allowing us to discuss one lobe at a time.  The black dots
mark the positions in phase space where the spectra of Fig.\ \ref{fig:sfdisp}
have been evaluated. The chemical potential difference $\delta\mu$ is measured
away from the lobe midpoint at $J=0$.}
\end{figure}

Jaksch {\it et al.}\cite{Jaksch98} pointed out that bosons in optical
lattices are accurately described by the Bose-Hubbard
Hamiltonian.\cite{fisher89} Depending on the value of the nearest-neighbor
hopping amplitude $J$, the on-site interaction $U$, and the chemical potential
$\mu$, the Bose-Hubbard model exhibits a superfluid or an insulating (with
$n_{0}\in{\mathbb N}$ particles per site) ground state, separated by a quantum
phase transition.\cite{greiner02} The phase diagram has been investigated on a
mean-field level,\cite{fisher89,oosten01} using perturbation
theory\cite{Elstner98} and with  numerical quantum Monte Carlo
methods.\cite{Batrouni90,Wessel04} Apart from one dimension, the qualitative
structure of the phase diagram is correctly described by mean-field
calculations, cf. Fig.~\ref{fig:lobes};  however, fluctuations tend to shift
the lobes to lower values of $J/U$ in two dimensions.\cite{Wessel04}

The dynamical properties of the Bose-Hubbard model were studied in both the
Mott and the superfluid  phase. In the Mott phase, the excitation spectrum was
studied numerically,\cite{Elstner98,Elstner99,Clark06} on a mean-field
level,\cite{oosten01} in a slave-particle approach,\cite{dickerscheid03,Lu05}
and in strong coupling perturbation theory.\cite{Sengupta04,Ohashi05} All
these approaches yield two gapped modes describing hole- and particle-type
branches.  For the weakly interacting superfluid ($J\gg U$) the
Gross-Pitaevskii\cite{Gross61,Pitaevskii61} equation  and  Bogoliubov theory
\cite{Abrikosov63} produce reliable results for the sound mode. However, for
the superfluid close to localization the character of the spectrum is still
unclear. We will show below that this spectrum involves two distinct modes: (i)
a sound mode characterized by a combined phase and density modulation, and (ii)
a gapped mode describing exchange between condensate and noncondensate at
fixed overall density.

The excitations of strongly correlated lattice bosons have been probed in two
experiments. In the experiment by Greiner {\it et al.}\cite{greiner02} the
whole lattice has been subjected to a global tilt. As this perturbation spoils
the translation invariance, it is difficult to infer bulk properties from
their results; the observed resonant response has been discussed in
Ref.~\onlinecite{Sachdev02}. Second, in their recent experiment, St\"oferle
{\it et al.}, \cite{Stoferle04} determine the energy absorption due to a
modulation of the lattice depth. They find a gapped continuum of excitations
for small $J/U$ and a broad feature in the superfluid. The appropriate
response function has been calculated in the framework of the Gross-Pitaevskii
equation\cite{Tozzo05} and for one-dimensional systems (1D), both
numerically\cite{Kollath06,Clark06} and analytically.\cite{Iucci05,Caux06} A
more powerful method to investigate this system is Bragg
spectroscopy,\cite{Stenger99} described by the dynamic structure factor
$S({\bf q},\omega)$. As compared to the above lattice modulation technique,
Bragg spectroscopy allows for a nonzero momentum transfer and hence reveals
the full structure of the excitation spectrum. However, the nonzero momentum
transfer requires to apply two additional lasers at a finite angle to the
system; so far, the limited optical access has hindered an experiment using
Bragg spectroscopy on strongly correlated lattice bosons.  On the theory side,
the structure factor $S({\bf q}, \omega)$ has been calculated in mean-field
theory\cite{Oosten05} and for 1D systems using numerical methods.\cite{Batrouni05,Reischl05} Below, we calculate both response functions
within the Mott insulator as well as in the strongly correlated superfluid,
making use of one unified description.

Our method is based on a technique previously introduced by Altman and
Auerbach\cite{Altman02} describing the particle-hole symmetric limit of the
Bose-Hubbard model at large particle numbers $n_0 \gg 1$. Here, we generalize
this approach to deal with the experimentally relevant regime of Mott
insulators with one and two particles per lattice site.  The method involves a
truncation of the Hilbert space to three states per site and a spin-wave
technique within a slave-boson language to describe fluctuations above a
mean-field ground state.  The truncation limits the validity of our results to
a single Mott lobe and its surrounding superfluid environment.

In the following Sec.~\ref{chap:vmf}, we introduce the variational mean-field
method outlined above. After proper truncation of the Hilbert space, we
construct the mean-field phase diagram using a variational ground state. We
proceed with the derivation of an effective Hamiltonian describing residual
particle fluctuations using a method motivated by the mapping to a spin-1
Hamiltonian.  We diagonalize this effective Hamiltonian with the help of a
generalized Bogoliubov transformation and find the spectra in both insulating
and superfluid phases. Section~\ref{chap:lrt} is devoted to the study of the
response functions: we discuss the dynamic structure factor (density-density
correlations) in the Mott phase as well as in the superfluid and compare our
findings with previously obtained theoretical results. In addition to Bragg
spectroscopy, we analyze the lattice modulation technique which is described
in terms of a dynamic modulation of the tunneling amplitude $J$ and calculate
the corresponding response function (hopping correlator) in both phases and
for varying dimensionality of the excitation. We summarize and conclude our
work in Sec.  \ref{chap:summary}.

%%%%%%%%%%%%%%%%%%%%%%%%%%%%%%%%%%%%%%%%%
\section{Variational mean-field}
\label{chap:vmf}

\subsection{Method}

The Bose-Hubbard Hamiltonian in a notation suitable for our approximation
scheme takes the form
\begin{equation}
\label{eqn:bhh}
H_{\rm BH} = 
-J \sum_{\langle i,j \rangle}a_{i}^{\dag}a_{j} 
+ \frac{U}{2} \sum_{i} \delta n_{i}^{2}
-\delta\mu \sum_{i} n_{i},
\end{equation}
where $a_{i}^{\dag}$ is the bosonic creation operator for a Wannier state at
site $i$, $n_{i}=a^{\dag}_{i}a_{i}$ is the number operator and $\delta
n_{i}=n_{i}-n_{0}$ measures deviations of the particle number from a mean
filling $n_{0}$. The chemical potential $\delta\mu$ is measured from the
middle of the lobe (cf. Fig.~\ref{fig:lobes}).

Our goal is the determination of the dynamical properties of the Bose-Hubbard
model in the limit of strong interactions. In particular, we are interested in
finding the excitation spectra and eigenstates in the Mott-insulating as well
as in the superfluid phase nearby.  Note that weakly interacting theories such
as the Gross-Pitaevskii equation  or the Bogoliubov theory cannot capture the
physics close to localization. On the other hand, strong coupling perturbative
approaches are often incapable to correctly describe the broken ${\sf
U}(1)$-symmetry phase.\cite{Sengupta04} Altman and Auerbach \cite{Altman02}
introduced a Hilbert space truncation method for large filling $n_{0}\gg1$,
i.e., the particle-hole symmetric case, which we extend to low fillings
where particle-hole symmetry is broken.  The basic idea is to truncate the
bosonic Fock space to only three local states. In this truncated space, we
first find a variational (mean-field) ground state and then derive an
effective Hamiltonian $H_{\rm eff}$ for the excitations above this ground
state.

The truncation to three local states with particle numbers $n_{0} $ and
$n_{0}\pm1$ is motivated by the strong suppression of particle number
fluctuations in, and close to, the Mott phase; its validity is discussed in
Sec.~\ref{chap:valid}, below. We introduce bosonic operators that create
`particles' in the retained three states
\begin{align}
\label{eqn:tdefn}
\nonumber
t_{1,i}^{\dag}|{\rm vac}\rangle
&= \frac{(a_{i}^{\dag})^{n_{0}+1}}{\sqrt{(n_{0}+1)!}}|{\rm vac}\rangle, \\
t_{0,i}^{\dag}|{\rm vac}\rangle
&= \frac{(a_{i}^{\dag})^{n_{0}}}{\sqrt{n_{0}!}}|{\rm vac}\rangle, \\
\nonumber
t_{-1,i}^{\dag}|{\rm vac}\rangle
&= \frac{(a_{i}^{\dag})^{n_{0}-1}}{\sqrt{(n_{0}-1)!}}|{\rm vac}\rangle,
\end{align}
where $|{\rm vac}\rangle$ denotes the state with no particles present. The
original bosonic operators $a_{i}$ can be expressed in terms of the
$t_{\alpha,i}$-operators ($\alpha=-1,0,1$),
\begin{equation}
\label{eqn:aoft}
a_{i}^{\dag}=\sqrt{n_{0}+1}t_{1,i}^{\dag}t^{}_{0,i}+
\sqrt{n_{0}}t_{0,i}^{\dag}t^{}_{-1,i}.
\end{equation}
The Hilbert space spanned by the $t_{\alpha,i}$-operators is too large and the
physical subspace is obtained by imposing the constraint
\begin{equation}
\label{eqn:con_t}
\sum_{\alpha=-1}^{1} t_{\alpha,i}^{\dag}t_{\alpha,i}=\openone.
\end{equation}

The possibility to map the above truncated bosonic problem to a spin-1
Hamiltonian (see Appendix~\ref{chap:spinhamiltonian}) motivates a strategy
inspired by the spin-wave theory above a ferromagnetic or antiferromagnetic ground state:\cite{Auerbach94} starting out by expressing the spin operators $S_\pm$ and $S_z$ via Schwinger bosons $a_{\scriptscriptstyle\rm SB}^{\dag}$,
$b_{\scriptscriptstyle\rm SB}^{\dag}$: $S_+ = a_{\scriptscriptstyle\rm
SB}^{\dag} b_{\scriptscriptstyle\rm SB}$, $S_- = b_{\scriptscriptstyle\rm
SB}^{\dag} a_{\scriptscriptstyle\rm SB}$, and $S_z = (
a_{\scriptscriptstyle\rm SB}^{\dag} a_{\scriptscriptstyle\rm SB} -
b_{\scriptscriptstyle\rm SB}^{\dag} b_{\scriptscriptstyle\rm SB})/2$, the
constraint $a_{\scriptscriptstyle\rm SB}^{\dag} a_{\scriptscriptstyle\rm SB} +
b_{\scriptscriptstyle\rm SB}^{\dag} b_{\scriptscriptstyle\rm SB} = 2S$ is used
in the ordered phase to go over to Holstein-Primakoff bosons
$b_{\scriptscriptstyle\rm SB}^{\dag} \to b_{\scriptscriptstyle\rm HP}^{\dag}$
and $a_{\scriptscriptstyle\rm SB}^{\dag} \to
\sqrt{\smash[b]{2S-b_{\scriptscriptstyle\rm HP}^{\dag}
b_{\scriptscriptstyle\rm HP}}}$, with subsequent expansion of the square root
in $b_{\scriptscriptstyle\rm HP}^{\dag} b_{\scriptscriptstyle\rm HP}$. In
order to realize this program in the present situation, we first have to find
the ground state playing the role of the ordered state in the spin problem. In
a second step, we implement the holonomic constraint (\ref{eqn:con_t}) via a
procedure analogous to the change from Schwinger to Holstein-Primakoff
bosons.

In order to find a proper ground state of the truncated problem we introduce
the following variational operators
\begin{align}
\nonumber
b_{0,i}^{\dag}& =  \cos(\vartheta/2) t_{0,i}^{\dag} +
\sin(\vartheta/2) [\cos(\chi) t_{1,i}^{\dag}+
\sin(\chi)t_{-1,i}^{\dag}], \\
\nonumber
\label{eqn:vargs}
b_{1,i}^{\dag} & = -\sin(\vartheta/2) t_{0,i}^{\dag} +
\cos(\vartheta/2) [\cos(\chi) t_{1,i}^{\dag}+
\sin(\chi) t_{-1,i}^{\dag}], \\ 
b_{2,i}^{\dag} &= -\sin(\chi) t_{1,i}^{\dag}+\cos(\chi) t_{-1,i}^{\dag},
\end{align}
where the Gutzwiller-type ground state shall be given by
\begin{equation}
\label{eqn:groundstate}
|\Psi(\vartheta,\chi)\rangle = \prod_{i}b_{0,i}^{\dag}|{\rm vac}\rangle.
\end{equation}
The parameter $\vartheta$ controls the admixture of particle number
fluctuations in the ground state, whereas a deviation from integer filling is
accounted for by a nonvanishing $\sigma = \pi/4-\chi$. As the transformation
(\ref{eqn:vargs}) is unitary, the $b_{m,i}$ operators ($m=0,1,2$) obey the
constraint
\begin{equation}
\label{eqn:constraint}
\sum_{m=0}^{2} b_{m,i}^{\dag}b_{m,i}=\openone,
\end{equation}
cf. (\ref{eqn:con_t}). The relevant energy scale in the Mott phase is given by
the interaction strength $U$. For the discussion of the phase diagram all
energies are rescaled and denoted with a bar, e.g., $\bar J=J/U$.
Combining the Hamiltonian (\ref{eqn:bhh}) and the ansatz
(\ref{eqn:groundstate}) provides us with the variational energy per lattice
site $\bar \varepsilon_{\rm var}(\vartheta,\sigma) =  \langle
\Psi(\vartheta,\sigma)| \bar H_{\rm BH} |\Psi(\vartheta,\sigma)\rangle/N$
which is given by
\begin{multline}
\label{eqn:varenergy}
\bar\varepsilon_{\rm var}(\vartheta,\sigma) 
= \sin(\vartheta/2)^{2}\left[ \frac{1}{2}
-\delta\bar\mu \sin(2 \sigma)\right] -\frac{\bar Jz}{4} 
\sin^{2}(\vartheta)  \\
\times \left[n_{0} +\sqrt{(n_{0}+1)n_{0}}\cos(2\sigma)+
\frac{1}{2}(1+\sin(2\sigma))\right],
\end{multline}
where $z=2d$ is the coordination number and $N$ denotes the number of sites.
In the Mott-insulating phase, particle number fluctuations are absent within a
mean-field approximation and thus $\vartheta=0$; the parameter $\sigma$ then
drops out of the variational energy $\bar\varepsilon_{\rm var}
(\vartheta,\sigma)$ and can be set to zero. In the superfluid case, $\vartheta
\neq 0$ turns out to be a convenient order parameter and hence $\sigma$ is
eliminated via minimization of the variational energy (\ref{eqn:varenergy})
with respect to $\sigma$,
\begin{equation}
\label{eqn:sigma}
\sigma(\vartheta)=\frac{1}{2}\arctan\left(\frac{4 \delta\bar \mu +
\bar J z [\cos (\vartheta )+1]}{2\bar J z [\cos (\vartheta )+1] 
\sqrt{n (n+1)}}\right).
\end{equation}
This allows us to write $\bar \varepsilon_{\rm var}(\vartheta,\sigma)$ as a
function of $\vartheta$ alone. Within a Ginzburg-Landau treatment of the phase
transition, we reexpress $\vartheta$ in terms of the superfluid order
parameter parameter $\psi=\langle \Psi(\vartheta,       \sigma)| a_{i}
|\Psi(\vartheta,\sigma)\rangle/N = \sin(\vartheta)
[\sqrt{n_{0}+1}\cos(\pi/4\!-\!\sigma) + \sqrt{n_{0}}
\sin(\pi/4\!-\!\sigma)]/2$ and expand the variational energy
(\ref{eqn:varenergy}) in $\psi$,
\begin{equation}
\bar \varepsilon_{\rm var}(\vartheta,\sigma(\vartheta)) 
\approx \bar a(\bar J,\delta\bar \mu) \psi^{2} 
+\frac{ \bar b(\bar J,\delta\bar\mu)}{2} \psi^{4}.
\end{equation}
The calculation of the coefficients $\bar a(\bar J,\delta\bar\mu)$ and  $\bar
b(\bar J,\delta\bar\mu)$ is straightforward.  The sign change of $\bar a(\bar
J,\delta\bar\mu)$ marks the phase boundary and the roots of $\bar a(\bar
J,\delta\bar\mu)=0$ provide us with the known mean-field lobes
\begin{equation}
\delta\bar\mu_{\rm c}^{\pm}(\bar J)
=\frac{1}{2}
\left[
-\bar Jz\pm \sqrt{1-2\bar Jz(1\!+\!2n_{0})
+(\bar Jz)^{2}}
\right],
\end{equation}
cf. Fig.~\ref{fig:lobes}. The tip of the lobes can be found by equating
$\delta\bar\mu_{\rm c}^{+}=\delta\bar\mu_{\rm c}^{-}$, providing the critical
hopping $ z\bar J_{\rm c}=1/(\sqrt{n_{0}+1}+\sqrt{n_{0}})^{2}$.  Due to
particle-hole asymmetry, the line of integer density ($\sigma=0$) is bending
down according to (cf. Fig.~\ref{fig:lobes})
\begin{equation}
\label{eqn:deltamu}
\delta\bar\mu_{n_{0}}
=-\frac{1}{4}[\bar Jz+(\sqrt{n_{0}+1}+\sqrt{n_{0}})^{-2}];
\end{equation}
we refer to this line as the particle-hole symmetric line, which starts out
from the tip of the lobe $\delta\bar\mu^{\pm}_{\rm c}(\bar J_{\rm c})$ as
expected.

The above determination of the ground state (\ref{eqn:groundstate}) has
provided us with the phase diagram of the Bose Hubbard model and allows us to
proceed with the second step of our program, the implementation of the
constraint (\ref{eqn:constraint}) by going over to Holstein-Primakoff-type
bosons; thereby, the operator $b_{0,i}^\dagger$ plays the role of the
Schwinger boson $a_{\rm\scriptscriptstyle SB}^\dagger$ and the remaining
operators $b_{1,i}^\dagger$, $b_{2,i}^\dagger$ generate the excitations above
this ground state, as does the operator $b_{\rm\scriptscriptstyle HP}^\dagger$
in the spin problem.  We then eliminate one slave boson ($b_{0,i}$) via the
constraint
\begin{equation}
\label{eqn:root}
b_{0,i} = \sqrt{1-n_{1,i}-n_{2,i}},
\end{equation}
where $n_{m,i}=b_{m,i}^{\dag}b_{m,i}$ ($m=1,2$). Having chosen a good
`classical' ground state with potentially small fluctuations, we can expand
the square root in (\ref{eqn:root})
\begin{equation}
\label{eqn:rootexpansion}
b_{0,i} = \sqrt{1-n_{1,i}-n_{2,i}} \approx
(1-{\scriptstyle\frac{1}{2}}n_{1,i}-{\scriptstyle\frac{1}{2}}n_{2,i});
\end{equation}
the validity of the expansion will be discussed later (cf.\ Sec.\
\ref{chap:valid}).

We express the Hamiltonian (\ref{eqn:bhh}) in terms of the $b$ bosons and
eliminate the $b_{0,i}$ with (\ref{eqn:rootexpansion}). Collecting all terms
up to quadratic order in the $b_{m,i}$ provides the effective Hamiltonian
\begin{equation}
\label{eqn:effh}
H_{\rm eff}= J z \sum_{{\bf k} \in K} \vec{b}_{\bf k}^{\,\dag}
\left(
\begin{array}{cccc}
g^{\scriptscriptstyle -1}_{\scriptscriptstyle 11,{\bf k}} & 
g^{\scriptscriptstyle -1}_{\scriptscriptstyle 12,{\bf k}} & 
f^{\scriptscriptstyle -1}_{\scriptscriptstyle 11,{\bf k}} &
f^{\scriptscriptstyle -1}_{\scriptscriptstyle 12,{\bf k}}  \\
g^{\scriptscriptstyle -1}_{\scriptscriptstyle 21,{\bf k}} &
g^{\scriptscriptstyle -1}_{\scriptscriptstyle 22,{\bf k}} & 
f^{\scriptscriptstyle -1}_{\scriptscriptstyle 21,{\bf k}} & 
f^{\scriptscriptstyle -1}_{\scriptscriptstyle 22,{\bf k}}  \\
f^{\scriptscriptstyle -1}_{\scriptscriptstyle 11,{\bf k}} & 
f^{\scriptscriptstyle -1}_{\scriptscriptstyle 12,{\bf k}} & 
g^{\scriptscriptstyle -1}_{\scriptscriptstyle 11,{\bf k}} & 
g^{\scriptscriptstyle -1}_{\scriptscriptstyle 12,{\bf k}} \\
f^{\scriptscriptstyle -1}_{\scriptscriptstyle 21,{\bf k}} &
f^{\scriptscriptstyle -1}_{\scriptscriptstyle 22,{\bf k}} &
g^{\scriptscriptstyle -1}_{\scriptscriptstyle 21,{\bf k}} &
g^{\scriptscriptstyle -1}_{\scriptscriptstyle 22,{\bf k}}
\end{array}
\right)
\vec{b}_{\bf k},
\end{equation}
where $\vec{b}_{\bf k}=(b_{1,{\bf k}}, b_{2,{\bf k}}, b_{1,-{\bf k}}^{\dag},
b_{2,-{\bf k}}^{\dag})^{\rm T}$ and $K$ denotes the first Brillouin zone. We
have dropped a constant term to be discussed later. The coefficients of the
normal $g^{\scriptscriptstyle -1}_{\scriptscriptstyle rs,{\bf k}}$ and
anomalous $f^{\scriptscriptstyle -1}_{\scriptscriptstyle rs,{\bf k}}$ terms
are given in Appendix~\ref{chap:effective}. The appearance of anomalous terms in
the effective Hamiltonian (\ref{eqn:effh}) is analogous to the situation where
the corresponding spin problem is characterized by an antiferromagnetic
ground state. These anomalous terms are removed via a Bogoliubov
transformation which hybridizes creation and annihilation operators, thereby
generating a new ground state carrying particle number fluctuations. This is
in contrast to the ferromagnetic case, where the inclusion of quantum
fluctuations does not impact on the classical ground state.  The presence of
the anomalous terms $f^{\scriptscriptstyle -1}_{\scriptscriptstyle rs,{\bf
k}}$ away from $J=0$ (cf.\ Appendix~\ref{chap:effective}), shows that the
Mott-insulating state is `nonclassical' and carries fluctuations for all
finite values of $J$.

\subsection{Diagonalization procedures}

The diagonalization of the effective Hamiltonian (\ref{eqn:effh}) can be
achieved via a (real) Bogoliubov transformation
\begin{equation}
\label{eqn:bogoliubov}
M \vec\beta_{\bf k} 
= \vec b_{\bf k} \quad \mbox{with} \quad D = M^{\rm T}H_{\rm eff}{M}
\end{equation}
diagonal, where $\vec \beta_{\bf k}=(\beta_{x,{\bf k}},\beta_{y,{\bf k}},\beta_{ x,-{\bf k}}^{\dag},\beta_{y, -{\bf k}}^{\dag})^{\rm T}$. In the Mott phase $x\!=\!p$ ($y\!=\!h$) stands for particle and hole excitations respectively,
whereas in the superfluid phase these indices stand for sound ($x\!=\!s$) and massive ($y\!=\!m$) modes. Those bosonic
commutation relations that are not automatically fulfilled are imposed by the
additional condition
\begin{equation}
\label{eqn:condition}
M\Sigma M^{\rm T}=\Sigma,
\end{equation}
where the matrix $\Sigma$ is given by the outer product
\begin{equation*}
\Sigma\equiv
\left(
\vec b_{\bf k}^{} \vec b_{\bf k}^{\,\dag}
\right)^{\rm T}
-
\left(
\vec b_{\bf k}^{\, \dag}
\right)^{\rm T}\left(
\vec b_{\bf k}^{}
\right)^{\rm T}
={\rm diag}(1,1,-1,-1),
\end{equation*}
reminiscent of the metric tensor in Minkowski space. The group ${\sf O}(2,2)$,
given by all real $4\!\times\!4$-matrices $M$ fulfilling
(\ref{eqn:condition}), then shares many properties with the Lorentz group
${\sf O}(1,3)$, namely its decomposition in terms of `boosts' (transformations
of coordinates with different signs in the metric) and `rotations'
(transformations in a sector of the metric with equal signs). It turns out
that this decomposition provides a useful strategy for the diagonalization of
$H_\mathrm{eff}$ in the Mott phase, where the symmetries of $H_\mathrm{eff}$
allow for an efficient determination of the corresponding rapidities and
angles.  On the other hand, in the superfluid phase, these symmetries are
absent and the diagonalization of $H_\mathrm{eff}$ is preferably done by
mapping (\ref{eqn:bogoliubov}) and (\ref{eqn:condition}) to a non-Hermitian
eigenvalue problem.\cite{Avery74}
\begin{figure*}
\begin{center}
\includegraphics{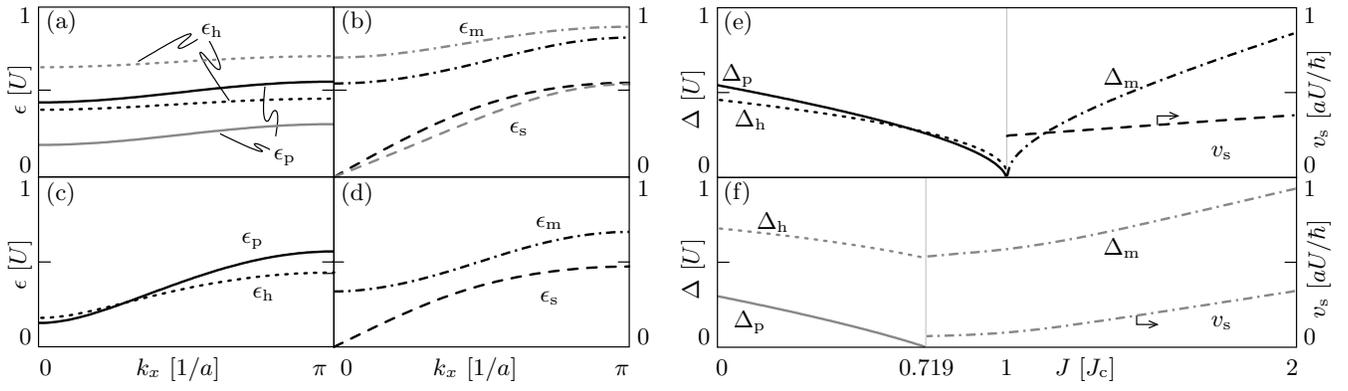}
\end{center}
\caption{
\label{fig:sfdisp}
The panels (a)-(d) show the spectra for different points in the phase
diagram (cf.\ Fig. \ref{fig:lobes}) in the $n_{0}=1$ lobe; all energies are
measured in units of $U$. We display the results for the two-dimensional case
with the dispersion along the direction $k_{x}$. Panel (a) and (c) refer to
the Mott phase, with the full line corresponding to the particle branch and
the dashed line to the hole branch. (a) $J=J_{\rm c}/3$; black
lines display dispersions on the line $\delta\mu_{n_{0}}$, gray lines 
correspond to $\delta\mu=0.2\, U$; the change in chemical potential produces
a shift in the spectra, cf.\ Eq.\ (\ref{eqn:mottexciations}). 
(c) $J=0.9\, J_{\rm c}$ and $\delta\mu=\delta\mu_{n_{0}}$; note the reduction
in the gaps $\Delta_{p(h)}$ as compared to (a).
Panels (b) and (d) refer to the superfluid phase with sound- and massive 
modes given by dashed and dash-dotted lines, respectively.
(b) $J=1.5\, J_{\rm c}$; black and gray lines refer to the 
chemical potentials $\delta\mu=\delta\mu_{n_{0}}$ and $\delta\mu=0.2\, U$. 
(d) $J=1.2\, J_{\rm c}$ closer to the transition and with $\delta\mu
=\delta\mu_{n_{0}}$. Panels (e) and (f) give the gap values
$\Delta_{p}$, $\Delta_{h}$, $\Delta_{m}$ and the sound velocity
$v_{s}$ for $J\in [0, 2J_{\rm c}]$ and $\delta\mu=\delta\mu_{n_{0}}$ 
(e) and $\delta\mu=0.2\, U$ (f). The value $J\approx 0.719\, J_{\rm c}$ 
corresponds to the phase boundary for $\delta\mu=0.2\, U$ 
(cf.\ Fig.\ \ref{fig:lobes}).}
\end{figure*}

{\bf (I)} In the  Mott state ($\sigma\!=\!0$, $\vartheta\!=\!0$), where no
anomalous mixing terms $f^{\scriptscriptstyle -1}_{\scriptscriptstyle 12,{\bf
k}}=f^{\scriptscriptstyle -1}_{\scriptscriptstyle 21,{\bf k}}=0$ between the
$b_{1,{\bf k}}$ and the $b_{2,{\bf k}}$ bosons are present (see
Appendix~\ref{chap:effective}), the parametrization of $M$ in terms of boosts and
rotations is suitable. To eliminate the anomalous terms $f^{\scriptscriptstyle
-1}_{\scriptscriptstyle 11,{\bf k}}$ and $f^{\scriptscriptstyle
-1}_{\scriptscriptstyle 22,{\bf k}}$ one chooses a boost in the $b^{}_{m,{\bf
k}}$-$b_{m,{\bf k}}^{\dag}$ plane; a subsequent rotation in the  $b_{1,{\bf
k}}$-$b_{2,{\bf k}}$ plane leads to the mean-field dispersions\cite{oosten01}
\begin{equation}
\label{eqn:mottexciations}
\epsilon_{p(h)}({\bf k}) 
= \mp[\epsilon_{0}({\bf k})/2 +\delta  \mu] +\tilde\omega({\bf k}),
\end{equation}
where $\tilde\omega({\bf k})=\sqrt{U^{2} - U \epsilon_{0} ({\bf k})  (4
n_{0}+2)+ \epsilon_{0}^{2} ({\bf k})}/2$ and $\epsilon_{0}({\bf k}) = 2 J
\sum_{l=1}^{d} \cos({\bf k}\cdot {\bf a}_{l})$ is the bare band dispersion.
Here, ${\bf a}_{l}$ denote the vectors connecting nearest neighbors and we
assume square and/or cubic symmetry, $a=|{\bf a}_{l}|$. The dispersions shown in
Figs.\ \ref{fig:sfdisp}(a) and (c) characterize two modes, describing particle
and hole type excitations, both with nonvanishing gaps $\Delta_{p (h)}
=\epsilon_{p (h)}(0)$, cf.\ Figs.\ \ref{fig:sfdisp}(e) and (f).  In the
Mott phase, the relation between the $t_{\alpha,{\bf k}}$ and the $b_{m,{\bf
k}}$ operators is trivial and the rotation in the $b_{1,{\bf k}}$-$b_{2,{\bf
k}}$ plane takes us back to the $t_{\alpha,{\bf k}}$ operators. We therefore
write the eigenstates in terms of the latter,
\begin{eqnarray}
\beta_{p,\bf k}^{\dag} &=& 
A({\bf k}) t_{1,{\bf k}}^{\dag}
+B({\bf k}) t_{-1,-{\bf k}}, \\
\beta_{h,\bf k}^{\dag}&=& 
-A({\bf k}) t_{-1,{\bf k}}^{\dag}
-B({\bf k}) t_{1,-{\bf k}},
\end{eqnarray}
with $A({\bf k})=\cosh({\rm arctanh}( f^{\scriptscriptstyle
-1}_{\scriptscriptstyle 11,{\bf k}}/g^{\scriptscriptstyle
-1}_{\scriptscriptstyle 11,{\bf k}})/2)$ and $B({\bf k}) =\sinh({\rm arctanh}(
f^{\scriptscriptstyle -1}_{\scriptscriptstyle 11,{\bf
k}}/g^{\scriptscriptstyle -1}_{\scriptscriptstyle 11,{\bf k}})/2)$, where we
have used the relations $g^{\scriptscriptstyle -1}_{\scriptscriptstyle 11,{\bf
k}}=g^{\scriptscriptstyle -1}_{\scriptscriptstyle 22,{\bf k}}$  and
$f^{\scriptscriptstyle -1}_{\scriptscriptstyle 11,{\bf
k}}=-f^{\scriptscriptstyle -1}_{\scriptscriptstyle 22,{\bf k}}$, which apply
for the Mott phase (cf. Appendix~\ref{chap:effective}).

{\bf (II)} In the superfluid phase, the coefficients $f^{\scriptscriptstyle
-1}_{\scriptscriptstyle 12,{\bf k}}$ do not vanish and furthermore,
$f^{\scriptscriptstyle -1}_{\scriptscriptstyle 11,{\bf k}} \neq
-f^{\scriptscriptstyle -1}_{\scriptscriptstyle 22,{\bf k}}$. The presence of
such terms renders a diagonalization via a parametrization of $M$ as in the
Mott state impractical. We therefore resort to the mapping onto a
non-Hermitian eigenvalue problem.\cite{Avery74} The constraint
(\ref{eqn:condition}) written as $M^{\rm T}=\Sigma M^{-1}\Sigma$ and inserted
into Eq.~(\ref{eqn:bogoliubov}) yields
\begin{equation}
\label{eqn:nonhermitian}
M^{-1} \Sigma H_{\rm eff} M = \Sigma D,
\end{equation}
The problem of finding a matrix $M\in {\sf O}(2,2)$ diagonalizing
$H_\mathrm{eff}$ is now shifted to the problem of diagonalizing the
non-Hermitian matrix $\Sigma H_{\rm eff}$.  The matrix $M$ then is obtained
from the eigenvectors $\{\tilde v\}$ of $\Sigma H_{\rm eff}$ via their proper
normalization (with respect to $\Sigma$): let $\tilde M$ be the matrix with
columns $\{\tilde v\}$; then $M \in {\sf O}(2,2)$ is given by
\[
M=L \tilde M \qquad \mbox{where} \qquad L=
{\rm diag}(l_{1},l_{2},l_{3},l_{4}),
\]
with
\[
l_{\alpha}^{-2} = \left(\tilde M \Sigma \tilde M^{\rm T}\right)_{\alpha
\alpha}
\]
(note that $\tilde M \Sigma \tilde M^{\rm T}$ is diagonal, i.e., the
eigenvectors $\{\tilde v\}$ are automatically orthogonal with respect to the
metric $\Sigma$).  After diagonalization the effective Hamiltonian reads
\begin{equation}
\label{eqn:soundmass}
H_{\rm eff} =\sum_{{\bf k} \in K}
\epsilon_{s}({\bf k})
\beta_{{s},{\bf k}}^{\dag}\beta_{{s},{\bf k}}^{}+
\epsilon_{m}({\bf k})
\beta_{{m},{\bf k}}^{\dag}\beta_{{m},{\bf k}}^{}
- C_{J,U}^{\delta\mu}.
\end{equation}
The optimization of the constant $C_{J,U}^{\delta\mu}$ leads to a
renormalization of $\vartheta$ and $\sigma$ and is discussed below. The
eigenvalues $\epsilon_{s(m)}({\bf k})$ can be calculated analytically (cf.
 Figs.~\ref{fig:sfdisp}(b)~and~(d)) and  we find a sound (Goldstone) mode,
which is linearly dispersing for $k\rightarrow 0$ with sound velocity $v_{\rm
s}\!=\!\partial_{k} \epsilon_{s}(k\!\!=\!\!0)/\hbar$ and a massive (Higgs)
mode with a gap $\Delta_{m}=\epsilon_{m}(0)$, cf.
Figs.~\ref{fig:sfdisp}(e)~and~(f). Note the vanishing of the particle- and hole
gaps and subsequent resurrection of the gap in the massive mode along the
particle-hole symmetric line; for $\delta \mu > \delta\mu_{n_0}$ the hole gap
transforms into the gap of the Higgs mode. The complete expressions for the
dispersions of the sound and massive modes turn out to be lengthy and are
given in Appendix~\ref{chap:effective}.

\begin{figure}[t]
\begin{center}
\includegraphics[width=\columnwidth]{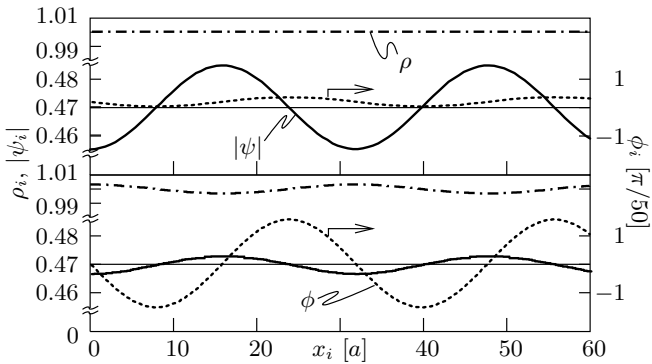}
\end{center}
\caption{
\label{fig:character}
Expectation values for the modulus $|\psi_{i}|$ and the phase
$\phi_{i}=\arg(\psi_{i})$ of the order parameter and the total density
$\rho_{i}$ for a coherent state of a massive mode (upper panel) and of a sound
mode (lower panel). We have chosen parameters $n_{0}=1$, $J=1.2 J_{\rm c}$,
$\delta\mu=\delta\mu_{n_{0}}$ guaranteeing particle-hole symmetry, resulting
in a ground-state order parameter $\psi_{0}\approx 0.47$ marked by the faint
gray line. Both modes refer to a ${\bf q}$ value along $x$ and $q\ll\pi/a$.
The massive mode is given by a local conversion of condensate and
noncondensate at fixed density, whereas the sound mode modulates the phase
with a corresponding modulation of the density.  In order to bring
$|\psi_{i}|$ ($\phi_{i}$) in both panels in registry we have chosen a phase
difference $\delta_{s,{\bf q}}-\delta_{m,{\bf q}}=\pi/2$.}
\end{figure}

To further characterize the excitations we consider coherent states of sound
[$|{\cal B}_{s,{\bf q}}\rangle$] and massive modes  [$|{\cal B}_{m,{\bf q}}\rangle$], respectively,
\begin{equation}
\label{eqn:coherent}
|{\cal B}_{s(m),{\bf q}}\rangle=
e^{-|{\cal B}_{s(m),{\bf q}}|^{2}/2}
e^{{\cal B}^{}_{s(m),{\bf q}}
\beta^{\dag}_{s(m),{\bf q}}}|0\rangle,
\end{equation}
where $|0\rangle$ denotes the vacuum with respect to the $\beta_{s(m),{\bf q}}$operators (i.e., the new ground state) and ${\cal B}_{s(m),{\bf q}}=|{\cal B}_{s(m),{\bf q}}|\exp[i\delta_{s(m),{\bf q}}]$ are complex numbers characterizing the coherent states. In
Fig.~\ref{fig:character} we present the expectation values of the density
operator $\rho_{i}=\langle {\cal B}_{s(m),{\bf q}}|a_{i}^{\dag}
a_{i}|{\cal B}_{s(m),{\bf q}}\rangle$ and the order parameter
$\psi_{i}=\langle {\cal B}_{s(m),{\bf q}}|a_{i}|{\cal B}_{s(m),{\bf q}}\rangle$ for $|{\cal B}_{s(m),{\bf q}}|\ll1$ and ${\bf q}$
along $x$ with $q\ll\pi/a$, i.e., we only add a small amount of long
wavelength excitations.  The sound mode is given by a modulation of the phase
accompanied by a modulation of the density, whereas the massive mode is given
by a local conversion of condensate and noncondensate. In an effective theory
for the order parameter $\psi_{i}$, the sound mode corresponds to the
Goldstone mode, whereas the massive mode corresponds to the Higgs boson. The
latter is only accessible for an effective theory which is second order in
time. In Ref. \onlinecite{Sachdev99}, the author develops such an action for
the superfluid to Mott-insulator phase transition.

The eigenvectors can be calculated analytically as well, however, for our
purpose the numerical solution of (\ref{eqn:nonhermitian}) is preferable.
Hermiticity allows us to write the transformation $M$ in the form
\begin{equation}
\label{eqn:sfelemts}
M({\bf k})=
\left(
\begin{array}{cc}
N({\bf k}) & P({\bf k}) \\
P(-{\bf k}) & N(-{\bf k})
\end{array}
\right);
\end{equation}
in addition, inversion symmetry renders the elements of the $2\!\times\!2$
matrices $N({\bf k})$ and $P({\bf k})$ independent of the sign of ${\bf k}$
and we can write
\begin{eqnarray}
\label{eqn:finalsftrans1}
\nonumber
b_{1,{\bf k}}^{\dag} & = & 
N_{11}({\bf k}) \beta_{{m},{\bf k}}^{\dag} + 
N_{12}({\bf k}) \beta_{{s},{\bf k}}^{\dag} +\\&&
P_{11}({\bf k}) \beta_{{m},-{\bf k}} + 
P_{12}({\bf k}) \beta_{{s},-{\bf k}}, \\
\label{eqn:finalsftrans2}
\nonumber
b_{2,{\bf k}}^{\dag} & = & 
N_{21}({\bf k}) \beta_{{m},{\bf k}}^{\dag} + 
N_{22}({\bf k}) \beta_{{s},{\bf k}}^{\dag} +\\&&
P_{21}({\bf k}) \beta_{{m},-{\bf k}} +  
P_{22}({\bf k}) \beta_{{s},-{\bf k}}.
\end{eqnarray}
The matrices $N({\bf k})$ and $P({\bf k})$ are calculated numerically and all
quantities of interest, e.g., response functions, are given in terms of
$N_{ij}({\bf k})$ and $P_{ij}({\bf k})$.

The constant term
\begin{equation}
\label{eqn:shift}
C_{J,U}^{\delta\mu} =Jz\sum_{{\bf k}\in K} 
g^{\scriptscriptstyle -1}_{\scriptscriptstyle 22,{\bf k}}+
g^{\scriptscriptstyle -1}_{\scriptscriptstyle 11,{\bf k}}
-(\epsilon_{s}({\bf k})+\epsilon_{m}({\bf k}))/2
\end{equation}
needs further discussion. The shift can be interpreted as a
fluctuation-induced reduction of the ground-state energy. In order to arrive
at a self-consistent description, the parameters $\vartheta$ and $\sigma$ have
to be determined by the new condition that the energy shift
$C_{J,U}^{\delta\mu}$ is maximal. This corresponds to finding a new density
and superfluid density for a given set of parameters $J,U,\delta\mu$. Only if
$\vartheta$ and $\sigma$ are renormalized with respect to their mean-field
values do we obtain a gapless sound mode, as demanded for the broken-symmetry
phase. In the Mott phase, $C_{J,U}^{\delta\mu}$ is not vanishing, but has its
maximum again at $\vartheta=\sigma=0$, thus generating no renormalization.

\subsection{Validity and expectation values}
\label{chap:valid}

The above method involves two approximations. First, the truncation of the
Hilbert space: the quality of the truncation is expected to be acceptable in
the Mott-insulating state and in the superfluid phase nearby, and turns bad in
the weakly interacting limit. The quality of the approximation can be checked
through comparison with experimental and numerical results, {\it e.g}.,
through testing the local number fluctuations. The latter have been measured
\cite{Greiner02a} and found to be suppressed due to strong interactions, at
least in the vicinity of the Mott phase. The data is consistent with 
subpoissonian number statistics as predicted in Ref.\ \onlinecite{Rokhsar91}
and supports the validity of the truncation to three local
states.\cite{Altman02} The technique allows for systematic improvement by the
inclusion of further local states.\cite{Defago06}

Second, while the expansion of the constraint (\ref{eqn:rootexpansion}) is not
{\it a priori} valid, we can check its quality via the ({\it a posteriori})
calculation of the ground state expectation values $\langle b_{m,i}^{\dag}
b_{m,i}^{} \rangle$: we have found values in the range from $0.18$ ($\bar J
\to \infty$) to $0.21$ ($\bar J = {\bar J}_{\rm c}$); fluctuations are largest
at the phase transition and go to zero in the Mott phase at $J\to0$ for
$n_{0}=1$ which justify our expansion in (\ref{eqn:rootexpansion}).

The calculation of matrix elements involves both states and the physical
operator in question: so far, we have calculated the spectrum within the
truncated Hilbert space, providing us with eigenvalues and eigenvectors of
$H_{\rm eff}$. In addition, we need to express the operators in the eigenbasis
generated by the $\beta$ operators. The specific step of replacing
`Schwinger bosons' by `Holstein-Primakoff' type operators involves the
elimination of all $b_{0,i}$ operators. Depending on the physical quantity
under consideration, its expression through the $\beta$ operators may or may
not involve an expansion of the square root (\ref{eqn:rootexpansion}); in
particular, the operators involving only $b_{1,i}$, $b_{2,i}$, and $n_{0,i} =
b_{0,1}^{\dag} b_{0,1}^{}$ (e.g., the density operator in the
Mott-insulating phase) can be transformed without requiring such an expansion.
Otherwise, an additional imprecision has to be accepted due to the square root
expansion of the constraint.

%%%%%%%%%%%%%%%%%%%%%%%%%%%%%%%%%%%%%%%%%
\section{Response functions}
\label{chap:lrt}

While cold atoms in optical lattices excel in their tunability, they do suffer
from a limited number of tools available for their characterization.  In fact,
so far only two experimental techniques are being used to determine the
dynamical properties of cold atoms in an optical lattice, Bragg spectroscopy
\cite{Stenger99} and lattice modulation \cite{Stoferle04}.  In Bragg
spectroscopy, two laser beams are focused on the system at an angle, leading
to an inelastic two-photon scattering process. The system's response is
described by the dynamic structure factor (density correlator)
\begin{equation}
\label{eqn:structurefactor}
S({\bf q},\omega)
=\sum_{n}\left| \langle n |\delta\rho_{-{\bf q}}|0\rangle\right|^{2}
\delta(\hbar \omega-\hbar\omega_{n0}),
\end{equation}
where $\delta\rho_{\bf q}$ is the density-fluctuation operator. The sum in
(\ref{eqn:structurefactor}) is running over all eigenstates $|n\rangle$ of the
system with $|0\rangle$ denoting the ground state and $\hbar\omega_{n0}$ the
excitation energy associated with $|n\rangle$. This method provides
angle-resolved information on the system, a feature which, however, turns out
to be responsible for the method's limitation in actual experiments, as the
optical access to the atom cloud is usually restricted.

In the lattice modulation technique, recently introduced by St\"oferle {\it et
al.}, the depth of the optical lattice is modulated with a frequency $\omega$,
 introducing side bands in the laser forming the optical lattice. Within the
framework of the Bose-Hubbard model (\ref{eqn:bhh}), the lattice modulation
corresponds to a modulation in the hopping parameter $J$; the determination of
the energy transfer then boils down to a calculation of the hopping correlator
\begin{equation}
\label{eqn:skin}
S^{\rm kin}_{(x)}(\omega)
=\sum_{n}\left| \langle n |T_{(x)}|0\rangle\right|^{2}
\delta(\hbar\omega-\hbar\omega_{n0}),
\end{equation}
with $T_{(x)}=\sum_{\langle i,j\rangle_{(x)}}a_i^{\dag}a_{j}$ the hopping
operator (the index $x$ refers to a restriction of the modulation along one
direction, here the $x$ axis). The energy absorption rate then is proportional
to $\omega S({\bf q},\omega)$ and $\omega S_{(x)}^{\rm kin}(\omega)$,
respectively. In order to evaluate the response functions
(\ref{eqn:structurefactor}) and (\ref{eqn:skin}) we have to express the
operators $\delta \rho_{\bf q}$ and $T_{(x)}$ in terms of the $\beta$ bosons,
involving a first transformation to $b$ bosons and subsequent elimination of
$b_{0,i}$, cf.\ Sec.\ \ref{chap:valid}.

We first concentrate on Bragg spectroscopy in the Mott and superfluid phases.
 In the Mott phase, where the local density is pinned to an integer value in
the ground state, we expect to excite a particle-hole continuum spread in
energy as described by the bandwidth of these two-particle excitations.
Surprisingly, we find pronounced peaks within this continuum which we can
relate to the single-mode excitations $\Delta_{h(p)}+\epsilon_{p(h)}({\bf k})$. In the superfluid phase one expects single-mode excitations,
as breaking the ${\sf U}(1)$ symmetry is leading to collective excitations,
and their weights will be determined.

Second, we proceed with the calculation of the hopping correlator
$S_{(x)}^{\rm kin}(\omega)$ in both the Mott-insulating and the superfluid
phase. Again we find a continuum in the insulating phase. In the superfluid
phase we do not expect to pump energy into the sound mode, as no momentum is
transferred with this probe (up to a reciprocal lattice vector).  A signal at
finite energy will then give direct access to the massive mode.

\subsection{Structure factor in the Mott phase}

\begin{figure}[t] 
\begin{center}
\includegraphics{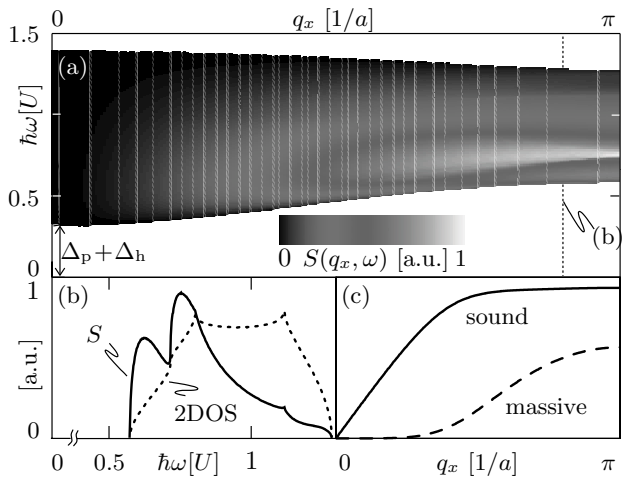} 
\end{center}
\caption{
\label{fig:skw_2d}
(a) Density plot of the dynamic structure factor $S({\bf q},\omega)$ in the
Mott phase along $q_{x}$ for $J=0.9J_{\rm c}$ [point (c) in Fig.\
\ref{fig:lobes}, energies are measured in units of $U$ and $n_{0}=1$, the
chemical potential $\delta\mu$ has no influence on the structure factor].
The spectrum is gapped, the spectrum's bandwidth decreases towards the
zone edge, while at the same time developing a pronounced structure
allowing for the identification of single-mode excitations.
(b) Cut along the $\omega$ direction at $q_{x}\!=\!0.9\pi/a$ [dotted
line in (a)]. The dynamic structure factor (solid line) exhibits
marked peaks at the single-mode energies; the dotted line shows the
2DOS exhibiting van Hove singularities. (c) Single-mode weights of
dynamic structure factor $S({\bf q},\omega)$ in the superfluid phase
at $J\!=\!1.2J_{\rm c}$ and $\delta\mu\!=\!\delta\mu_{n_{0}}$.
The sound mode (Goldstone, solid line) exhausts all the available
spectral weight at long wavelengths, allowing the massive mode (Higgs,
dashed line) to gain weight only towards the zone boundary ($[{\rm a.u.}]$
denotes `arbitrary units').}
\end{figure}

We make use of the eigenstates obtained within the variational mean-field
approach to calculate the dynamic structure factor (\ref{eqn:structurefactor})
in the Mott phase. In the truncated space the density-fluctuation operator
\begin{equation}
\label{eqn:generalrho}
\delta \rho_{\bf q} = 
\sum_{i} (a_{i}^{\dag}a_{i}-\langle a_{i}^{\dag}a_{i} \rangle)
e^{-i {\bf q}\cdot {\bf r}_{i}}
\end{equation}
takes the form
\begin{equation}
\label{eqn:densfluc-t}
\delta \rho_{\bf q} = \sum_{{\bf k} \in K} t_{1,{\bf k}}^{\dag}
t_{1,{\bf k}+{\bf q}}^{} - 
t_{-1,{\bf k}}^{\dag} t_{-1,{\bf k}+{\bf q}}^{},
\end{equation} 
where we have used the {\it exact} constraint (\ref{eqn:constraint}) to
eliminate $t_{0,i}$ ($=b_{0,i}$ in the Mott phase). Going over to
$\beta$ operators we obtain
\[
\delta \rho_{\bf q} = \!\! \sum_{{\bf k} \in K}
A({\bf k}) B({\bf k}+{\bf q})
\bigl[
\beta_{p,\bf k}^{\dag} 
\beta_{h,-({\bf k}+{\bf q})}^{\dag} -
\beta_{h,\bf k}^{\dag} 
\beta_{p,-({\bf k}+{\bf q})}^{\dag}
\bigr]
\]
and inserting this expression into the formula for the dynamic structure
factor yields
\begin{equation}
\label{eqn:skw-integral}
S({\bf q},\omega) = \frac{1}{2} \int_{K} \frac{d{\bf k}}{v_{0}}\, 
P({\bf k}, {\bf q}) \delta[\hbar \omega-\epsilon_{h}({\bf k})-
\epsilon_{p}({\bf q}-{\bf k})],
\end{equation}
with $v_{0}=(2\pi/a)^{d}$ the volume of the Brillouin zone.  The matrix
element $P({\bf k},{\bf q})$ quantifies the coupling of each particle-hole
excitation to the density perturbation and is given by
\begin{eqnarray}
\label{eqn:correlation}
&&1+P({\bf k},{\bf q}) = \\
\nonumber
&&\!\!\! \frac{\epsilon_{0}({\bf k}) \epsilon_{0}({\bf q-k})
+U^{2}-U(2n_{0}+1)
[\epsilon_{0}({\bf k})+\epsilon_{0}({\bf q-k})]}
{4\tilde\omega({\bf k}) \tilde\omega({\bf q-k})}.
\end{eqnarray}
The dynamic structure factor (\ref{eqn:skw-integral}) is closely related to
the two-particle density of states (2DOS) given by
\begin{equation}
\label{eqn:2DOS}
D({\bf q},\omega) = \int_{K} \frac{d{\bf k}}{v_{0}}\, 
\delta[\hbar\omega-\epsilon_{h}({\bf k})-
\epsilon_{p}({\bf q}-{\bf k})].
\end{equation}
They share the same bandwidth as well as the characteristic total gap  of the
insulating state
\begin{equation}
\label{eqn:total_gap}
\Delta_{\rm tot}
=\Delta_{p}+\Delta_{h}=\sqrt{U^{2}-2JzU(2n_{0}\!+\!1)+(Jz)^{2}}.
\end{equation}
Moreover, the van Hove singularities present in the 2DOS influence the
response function, see Fig.\ \ref{fig:skw_2d}(b).

In our discussion of the results,\cite{comment:3d} see Figs.
\ref{fig:skw_2d}(a) and (b), we concentrate on the two-dimensional situation
where our mean-field analysis is sufficiently accurate. In the Mott phase,
particle-hole excitations lead to a continuum, starting with a finite gap
$\Delta_{\rm tot}$ at zero momentum ${\bf q}$. The bandwidth decreases with
increasing momentum transfer following the support of the 2DOS and reaches a
minimum at the zone boundary, see Fig.\ \ref{fig:skw_2d}(a).  Figure
\ref{fig:skw_2d}(b) shows the 2DOS $D({\bf q},\omega)$ as well as $S({\bf
q},\omega)$ at ${\bf q}=(0.9\pi/a) {\bf e}_{x}$: The van Hove singularities in
the 2DOS are washed out in the response function which is dominated by the
matrix element $P({\bf k},{\bf q})$. The latter generates the two pronounced
peaks in $S({\bf q},\omega)$ which we find (numerically) to be located at
$\Delta_{p}+\epsilon_{h}(q_{x})$ and $\Delta_{h}+\epsilon_{\rm
p}(q_{x})$. Hence, although Bragg spectroscopy generically excites a
two-particle continuum, tracing these peaks allows for the extraction of the
single-particle energies. These peaks are most prominent near the zone
boundary and can be enhanced at small values of $q_x$ by dividing out a global
modulation of the form $[1\!-\!\cos(q_{x}/a)]$ (cf.\ Appendix~\ref{chap:rspt})
from $S({\bf q},\omega)$. Furthermore, the peaks disappear deeper in the Mott
phase where the excitations are more localized.  We note that by expanding the
integral (\ref{eqn:skw-integral}) in $J/U$, we recover the results of the
perturbative treatment (see Appendix~\ref{chap:rspt}), hence the quadratic
expansion of the constraint (\ref{eqn:constraint}) is consistent with
second-order perturbation theory; however, the two results are comparable only
for $J \lesssim 10^{-2} J_c$.

\subsection{Structure factor in the superfluid phase}

In the superfluid phase, the expression for the density fluctuation operator
(\ref{eqn:densfluc-t}) written in terms of the $b$ bosons contains $b_{0,i}$
operators, which have to be removed via an expansion of the square root
(\ref{eqn:rootexpansion}), leading to terms linear in the operators $b_{m,
{\bf k}}^{\dag}$ as well as higher-order terms. Going over to
$\beta$ operators we obtain the expression
\begin{multline}
\label{eqn:expandedrho}
\delta\rho_{\bf q}= 
\sin(\vartheta/2)
\Big\{ 
[P_{21}({\bf q})+N_{21}({\bf q})] \beta_{{m},{\bf q}}^{\dag}
\\
[P_{22}({\bf q})+N_{22}({\bf q})] \beta_{{s},{\bf q}}^{\dag}
\Big\}+O(\beta^{2}).
\end{multline}
Accounting only for terms up to first order, we ignore two- and multiparticle
excitations and we cannot expect to fulfill the $f$-sum rule exactly, see the
discussion in Sec.~\ref{chap:particlenumber} below. Inserting
(\ref{eqn:expandedrho}) into the expression for the dynamic structure factor
yields
\begin{eqnarray}
\nonumber
\frac{1}{N} S^{(1)}({\bf q},\omega) & = & \sin^{2}(\vartheta/2) 
\bigg[ 
\tilde S^{2}({\bf q}) \delta(\hbar\omega-\epsilon_{s}({\bf q})) \\
& & \qquad \quad
\label{eqn:sf-single} 
\!\!\!\!\!\!\!\!\!+\, \tilde M^{2}({\bf q}) 
\delta(\hbar\omega-\epsilon_{m}({\bf q}))
\bigg];
\end{eqnarray}
this result reveals the two collective modes (\ref{eqn:soundmass})
characterizing the superfluid phase. Their weights are given by $\tilde M({\bf
q})\!=\!P_{21}({\bf q})\!+\!N_{21}({\bf q})$ for the massive mode and $\tilde
S ({\bf q})\!=\!P_{22}({\bf q})\!+\!N_{22}({\bf q})$ for the sound mode and
are shown in  Fig.~\ref{fig:skw_2d}(c). The sound mode is dominating the
response at low momenta, with the massive mode acquiring weight only for
higher momenta, where the sound mode saturates; in fact, numerical analysis
confirms $\propto q^{4}$ dependence of $\tilde M^{2}({\bf q})$ at small ${\bf
q}$.

Besides the single mode contribution, $\delta\rho_{\bf q}$ also yields
two-particle continua involving the excitations $|{m},{\bf k};{s},{\bf
q}-{\bf k}\rangle$, $|{m},{\bf k};{m},{\bf q}-{\bf k}\rangle$ and
$|{s},{\bf k};{s},{\bf q}-{\bf k}\rangle$. Their weight is about three
orders of magnitude smaller than the single-particle contribution, however.

\subsection{Particle-number conservation}
\label{chap:particlenumber}

The dynamic structure factor is constrained by several sum rules deriving from
conservation laws. Gauge symmetry and thus particle-number conservation is
leading to the well known $f$-sum rule
\begin{equation}
\label{eqn:orig-fsum}
\int_{0}^{\infty}d\omega\,\omega S({\bf q},\omega)=\frac{Nq^{2}}{2m},
\end{equation}
which is modified in a one-band lattice description. The broken translational
symmetry is leading to a nonquadratic dispersion, which can be characterized
by a ${\bf k}$-dependent effective mass tensor $\hbar^{2}/m^{*}_{ij}({\bf k})
= \partial_{k_{i}}\partial_{k_{j}}\epsilon_{0}({\bf k})$. The $f$-sum
rule\cite{Batrouni05,Huber04} adapted to the presence of a lattice then takes
the form
\begin{eqnarray}
\label{eqn:fsumrule}
\lefteqn{\int_{0}^{\infty} d \omega\, \omega S({\bf q},\omega)}\\ 
\nonumber
&=&\!\! 
\frac{1}{2\hbar^2}\sum_{{\bf k} \in K}\{\epsilon_{0}({\bf k}+{\bf q})+ 
\epsilon_{0}({\bf k}-{\bf q})-2 \epsilon_{0}({\bf k})\}
\langle 0 |a_{\bf k}^{\dag} a_{\bf k} |0 \rangle \\
\nonumber
&\stackrel{q\rightarrow0}{\approx}& 
\sum_{ij} \frac{ q_{i} q_{j}}{2} 
\sum_{{\bf k} \in K} 
\frac{\langle 0 |a_{\bf k}^{\dag} a_{\bf k} |0 \rangle} 
{m^{*}_{ij}({\bf k})} \\
\nonumber
&&\qquad \Bigl(= \frac{N §q^{2}}{2m} 
\qquad \mbox{for}\qquad 
\frac{1}{m_{ij}^{*}({\bf k})}\equiv\frac{1}{m}\delta_{ij} \Bigr). 
\end{eqnarray} 
Unlike in translation-invariant systems the structure of the ground state
enters the $f$-sum rule via the nonuniversal prefactor
\begin{equation}
\label{eqn:prefac}
I_{ij}(J/U)=\sum_{{\bf k}\in K}
\frac{\langle 0|a_{\bf k}^{\dag}a_{\bf k}|0\rangle}
{m_{ij}^{*}({\bf k})}.
\end{equation}
The expression (\ref{eqn:fsumrule}) predicts $\propto q^2$ behavior at small
$q$, which is trivially fulfilled in the superfluid phase (combine the weight
$\tilde{S}^2 \propto q$, see Fig.\ \ref{fig:skw_2d}(c), with the linear
dispersion of the sound mode; the $\propto q^{4}$ dependence of $\tilde M^{2}$
does not contribute at small $q$) and can be easily verified in the Mott phase
via expansion of the matrix element $P({\bf k},{\bf q})$,
\[
P({\bf q},{\bf k})=P^{(2)}({\bf k})q^{2} + O(q^4).
\]
Unfortunately, our scheme does not allow for a precise calculation of the
prefactor (\ref{eqn:prefac}) and hence an exact self-consistency check (via
particle-number conservation) of our result is not possible.

The issue of number conservation has been raised in the work of van Oosten
{\it et al.}\cite{Oosten05}. Their field-theoretic calculation of the
structure factor did not reproduce the required $q^2$ behavior, which then has
been enforced through the use of Ward identities. However, it appears that the
Green's function\cite{oosten01} $G(i\omega_{n},{\bf k})$ used in this
calculation already violates number conservation in the Mott phase, i.e.
\[
\rho_{i}=\langle n_{i} \rangle = 
\frac{1}{\beta}\sum_{i\omega_{n}}
\int_{K} \frac{d{\bf k}}{v_{0}}
\, G(i\omega_{n},{\bf k}) \neq n_{0}.
\]
The application of Ward identities, although guaranteeing number conservation,
generates other defects in the structure factor, e.g., the appearance of
linear terms in $J$ spoiling the $J\!\rightarrow\!-J$ symmetry present in
bipartite lattice models.

\subsection{Lattice modulation in the Mott phase}

The lattice-depth modulation is a particle-number conserving probe and hence
produces only particle-hole excitations. While the lattice modulation has been
uniaxial so far, \cite{Stoferle04} here, we also discuss its extension to an
isotropic modulation (we discuss the case of equal modulation amplitudes but
allow for mutual phase-differences between the various directions).  Expanding
the constraint to second (i.e., leading) order, we obtain the response
function\cite{comment:isotropy}
\begin{equation}
\label{eqn:skinmott}
S^{\rm kin}_{(x)}(\omega) =
\frac{1}{2} \int_{K}\frac{d{\bf k}}{v_{0}}\, 
P^{\rm kin}_{(x)}({\bf k})
\delta[\hbar\omega-\epsilon_{h}({\bf k})-\epsilon_{p}(-{\bf k})],
\end{equation}
with
\begin{equation}
\label{eqn:pkin}
P^{\rm kin}_{(x)}({\bf k})=
2n_{0}(n_{0}+1)|\Sigma_{(x)}({\bf k})|^{2}
\left[\frac{U} {\tilde \omega({\bf k})}\right]^{2}.
\end{equation}
The interference generated by the different lattice modulations is encoded in
the sum
\begin{equation}
\label{eqn:interfernce}
\Sigma({\bf k})=2\sum_{l=1}^{d}e^{i\phi_{l}}\cos({\bf k}\cdot {\bf a}_{l});
\end{equation}
for the uniaxial modulation this reduces to
\begin{equation}
\label{eqn:cosine}
\Sigma_{x}({\bf k})=2 \cos({\bf k}\cdot{\bf a}_{x}).
\end{equation}
The relative phase $\phi_{l}$ between the different lattice modulations can
lead to interesting interference effects, see below.

In Figs.~\ref{fig:shake}(b) and (c) the results for isotropic and uniaxial
lattice modulation are shown for $\phi_{l}= 0$ and $d=2$. The bandwidth is
determined by the 2DOS (\ref{eqn:2DOS}) at zero momentum transfer ${\bf q}$.
Energy is transferred to the system only at frequencies $\omega$ above the gap
$\Delta_{\rm tot}/\hbar$, offering a simple way to determine the gap value. A
dramatic change is obtained when going from the uniaxial to the isotropic
modulation: the cusp at $\hbar\omega\!=\!U$ disappears and is replaced by a
zero in the absorption probability. As the matrix element $P^{\rm kin}({\bf
k})$ is non-negative, the response (\ref{eqn:skinmott}) only disappears if
$P^{\rm kin}({\bf k})\equiv 0$ for all ${\bf k}$-values on the line defined by
the $\delta$-function in the integral (\ref{eqn:skinmott}). At nonzero
$\phi_{l}$, a finite weight is assembled away from the points $(\pm \pi/2,\pm
\pi/2)$, leading to a finite response at $\omega\!=\!U$. However, for
$\phi_{l}\ll\pi$ an appreciable suppression is still observable.

The result (\ref{eqn:skinmott}) obtained here has to be compared with the one
obtained by Iucci and coworkers:\cite{Iucci05} in their perturbative
calculation the factor $[U/\tilde\omega({\bf k})]^{2}$ does not show up.
Instead, the Bogoliubov transformation used here is equivalent to a
resummation of diagrams and leads to this factor generating the interesting
structure in the response function (\ref{eqn:skinmott}).

Comparing our result with the experiment of St\"oferle {\it et al.}
\cite{Stoferle04} we have to consider the uniaxial case, i.e., $S^{\rm
kin}_{x}(\omega)$. While St\"oferle {\it et al.} observe a broad two-peak
structure with maxima around $U$ and $2U$, the present accuracy of our
calculation does not account for high-energy excitations residing around $2U$.
 On the other hand, the current experimental resolution does not allow to
trace the interesting structure on the scale of the bandwidth, see Figs.\
\ref{fig:shake}(b) and (c).  Thus, both theory and experiment have to be
developed further in order to allow for a precise comparison. Furthermore, it
is worth mentioning that the current experiment may probes non-linear
response.\cite{Kraemer04}
\begin{figure}[t]
\begin{center}
\includegraphics{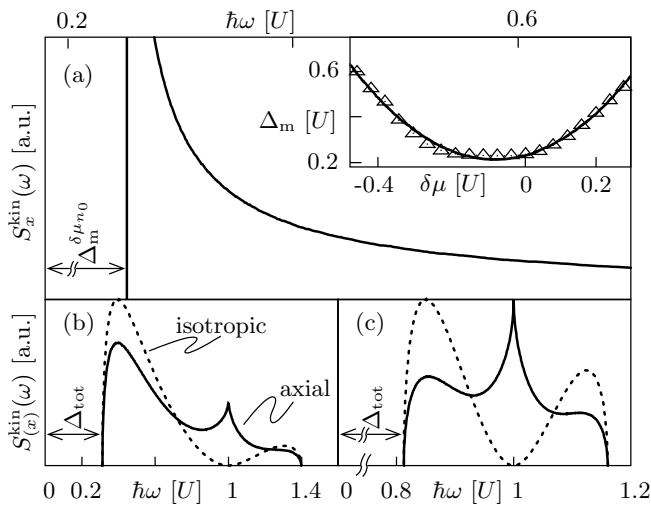}
\end{center}
\caption{
\label{fig:shake}
(a) Hopping correlator $S_{x}^{\rm kin}(\omega)$ in the superfluid phase for
$J=1.2J_{\rm c}$ and $n_{0}=1$ for a parabolic trap and commensurate filling
in the center (in a homogeneous system the response is nonzero only at
$\omega=\Delta_{m}$). The inset shows the gap $\Delta_{m}$ of the
massive mode as a function of the chemical potential with a minimum at
commensurate filling $\delta\mu = \delta\mu_{n_0}$.  The triangles are the
calculated gap values and the solid line is a fit used in the calculation of
$S^{\rm kin}_{x}(\omega)$ in a trap. (b) $S_{(x)}^{\rm kin}(x)$ in the Mott
phase at $J=J_{\rm c}/3$. The response consists of a particle-hole continuum
with a gap $\Delta_{\rm tot}$. The striking difference between the uniaxial
[$S_{x}^{\rm kin}(\omega)$, solid line] and the isotropic [$S^{\rm
kin}(\omega)$, dashed line] situation is due to interference effects. (c) The
same as in (d) for $J=0.9 J_{\rm c}$, where the gap $\Delta_{\rm tot}$ is
larger and the available bandwidth smaller.  }
\end{figure}

\subsection{Lattice modulation in the superfluid phase}

In the superfluid phase (with an order parameter $|\psi|>0$), the expansion of
the constraint again provides terms linear in the $\beta$ operators and we
obtain single mode peaks in the hopping correlator $S^{\rm kin}_{x}$ at
$\hbar\omega=0$ and $\hbar\omega=\Delta_\mathrm{m}$ due to sound and massive
excitations, respectively. Energy absorption is due to the excitation of the
massive mode only and its weight $W_{\rm mass} \propto |\psi|^2$ is given in
Appendix\ \ref{chap:effective}. In addition, we also find a two-mode continuum
(dominated by processes $|{s},{\bf k};{m}, -{\bf k}\rangle$) with a
weight suppressed by more than three orders of magnitude in the entire
parameter range.

In a trap, the sharp peak at $\Delta_\mathrm{m}$ will be smeared due to the
inhomogeneous density distribution. [Note that in the Mott phase, the response
is not changed by the trap as both excitation energies $\epsilon_{h}({\bf
k})+\epsilon_{p}({\bf q}-{\bf k})$ (\ref{eqn:mottexciations}) and the
matrix elements $P({\bf k},{\bf q})$ and $P^{\rm kin}_{(x)}({\bf k})$ are
independent of $\delta\mu$]. The onset of absorption is determined by the
minimal gap $\Delta_{m}^{\rm min}$ which occurs on the particle-hole
symmetric line $\delta\mu = \delta\mu_{n_0}$ and the shape of the absorption
profile depends on the distribution of effective chemical potentials
$\mu_\mathrm{eff}(r)=\mu- V_\mathrm{trap}(r)$ in the trap. Here, we analyze
the shape for a quadratic trap with $\mu_\mathrm{eff}(0)=\delta\mu_{n_0}$ for
a superfluid close to the Mott phase, $J\!=\!1.2 J_{\rm c}$. The dependence of
the gap $\Delta_{m}$ on the chemical potential $\delta\mu$ can be
calculated, see the inset of Fig.~\ref{fig:shake}(a), and a convenient fit is
given by $\Delta_{m}\!-\!\Delta_{m}^{\rm min} \!=\!w
(\delta\mu\!-\!\delta\mu_{n_{0}})^{2}$ with $w\approx 2.06U^{-1}$. The hopping
correlator then draws its weight at frequency $\omega$ from rings matching the
local gap energy $\Delta_\mathrm{m}(r)$,
\begin{eqnarray}
\nonumber
S_{x}^{\rm kin}(\omega)&=&\frac{1}{\pi R^2} \int_0^R dr\,2\pi r
W_\mathrm{mass} 
\delta[\hbar\omega-\Delta_\mathrm{m}(r)]\\
\label{eqn:deltatrap}
&\propto& \frac{1}{\sqrt{\hbar\omega-\Delta_{m}^{\rm min}}} \quad
\mbox{for} \quad \hbar\omega > \Delta_{m}^{\rm min}.
\end{eqnarray}
The resulting tail then resembles the broad absorption profile observed in the
experiment of St\"oferle {\it et al.}\cite{Stoferle04,Schori04} Note that the
precise shape depends on the actual density distribution in the trap; in
particular, the divergence at $\Delta_{m}^{\rm min}$ is removed when
$\mu_\mathrm{eff} (0)\neq \delta\mu_{n_0}$.  Another potential source of
broadening is the finite lifetime of the massive mode due to the decay into
two phonons as considered by Altman and Auerbach in Ref.\
\onlinecite{Altman02}. However, in two and more dimensions the effect of the
trap dominates over the lifetime broadening.

%%%%%%%%%%%%%%%%%%%%%%%%%%%%%%%%%%%%%%%%%
\section{Summary and Conclusions}
\label{chap:summary}

We have generalized the truncation scheme, introduced by Altman and Auerbach
\cite{Altman02} to deal with the Bose-Hubbard model in the particle-hole
symmetric limit (large particle numbers $n_0\gg 1$), to the experimentally
relevant situation of small filling numbers $n_0$ of order unity. The
determination of excitations consisting of small fluctuations about a
variational ground state is inspired by the Holstein-Primakoff description of
quantum-spin systems and corresponds to the determination of spin-wave
excitations above an antiferromagnetic ground
state.\cite{comment:dm}\nocite{Dyson56,Maleev58} We have determined the
mean-field phase diagram as well as the spectra and eigenstates in the
Mott-insulator and superfluid phases. These results then have been applied to
the calculation of two response functions, the structure factor (density
correlator) describing Bragg spectroscopy and the hopping correlator
describing the lattice-modulation spectroscopy.

A mean-field variational ansatz provides us with the usual phase diagram valid
in dimensions $d \geq 2$. The inclusion of two additional levels allows us to
account for particle and hole-type excitations in the Mott phase and we
determine, using a Bogoliubov transformation, the dispersions of the two
gapped modes; at the same time, the Bogoliubov transformation introduces a new
ground state carrying particle-hole fluctuations. In the superfluid phase, we
find a sound (Goldstone) and a massive (Higgs) mode and we determine the
characteristic velocity and gap parameters. The massive mode describes a
local counterflow of condensate and normal densities; this mode is absent in a
Gross-Pitaevskii description where the dynamics involves a first-order time
derivative, but is allowed in a Klein-Gordon type theory with a second-order
dynamics. The presence of the latter is due to the underlying Mott physics
providing `particles' and `anti-particles' (holes) and has been shown to be
relevant in the Mott-insulator--superfluid transition in Ref.\
\onlinecite{Sachdev02} (an additional first-order derivative is present away
from the particle-hole symmetric line). A similar (massive) mode was found for
the charge-density-wave compound NbSe$_{2}$ below the superconducting
transition temperature.\cite{Sooryakumar80,Lei85} The fate of this gapped mode
for $U\rightarrow0$, where higher occupation numbers are of importance, is
currently under investigation.\cite{Defago06}

In the Mott phase, Bragg spectroscopy excites a particle-hole continuum and
provides information on the gap and bandwidth of these two-particle
excitations. To our surprise, we find that the structure factor unveils the
single-particle excitation energies as well. In the superfluid phase,
collective modes (sound and massive) are excited and visible as sharp peaks
(to be smeared in a trap); the massive mode gains weight only at large
momenta. The transition can be traced watching the appearance of a gap when
crossing from the superfluid into the Mott-insulating phase.

The lattice-modulation scheme\cite{Stoferle04} is presently the tool of choice
to gain spectroscopic information on atomic matter in optical lattices. Our
calculation of the hopping correlator providing the system's response reveals
a two-particle continuum in the Mott phase which is sensitive to the details
of the excitation scheme, uniaxial versus isotropic. The response in the
superfluid is determined by the massive mode which appears as a sharp peak at
finite frequencies. In order to better account for the experimental results we
have extended our analysis to include the smearing due to the trap and find
that the precise shape depends sensitively on the value of the chemical
potential in the trap center. The experimental detection of such an energy
absorption at finite frequency cannot be easily understood within a weakly
interacting theory as described by the Gross-Pitaevskii equation.  On the
other hand, our strongly interacting theory provides a massive (Higgs) mode
which naturally accounts for such a finite frequency absorption. Furthermore,
a future experiment could address the question of how this Higgs mode
disappears in the weekly interacting regime.

We thank A.\ Castro-Neto, T.\ Esslinger, M.\ K\"ohl, C.\ Kollath, F.\ Hassler,
and A.\ R\"uegg for discussions and the Swiss NSF for financial support via
the NCCR MaNEP. The numerical work was performed on the Hreidar cluster at ETH
Z\"urich. Work in Innsbruck was supported by the Austrian Science Foundation,
and the Institute for Quantum Information.

%%%%%%%%%%%%%%%%%%%%%%%%%%%%%%%%%%%%%%%%%
\appendix

\section{Spin Hamiltonian}
\label{chap:spinhamiltonian}

Defining the spin-1 operators $S_{i}^{+}$, $S_{i}^{-}$ and $S_{i}^{z}$ in
terms of the $t_{\alpha,i}$ operators via
\begin{align}
S_{i}^{+}&=\sqrt{2}(t_{1,i}^{\dag}t_{0,i}+t_{0,i}^{\dag}t_{-1,i}) \\
S_{i}^{-}&=\sqrt{2}(t_{0,i}^{\dag}t_{1,i}+t_{-1,i}^{\dag}t_{0,i}) \\
S_{i}^{z}&=t_{1,i}^{\dag}t_{1,i}-t_{-1,i}^{\dag}t_{-1,i} 
\qquad\mbox{with} \\
[S_{i}^{+},S_{i}^{-}]&=2S_{i}^{z}, \quad [S_{i}^{z},S_{i}^{\pm}] =
\pm S_{i}^{\pm},
\end{align}
we can write the Bose Hubbard Hamiltonian (\ref{eqn:bhh}) in the truncated
space as
\begin{multline}
\label{eqn:spinhamiltonian}
H_{\rm BH}^{\rm spin}=
-\frac{ J n_{0}}{2} \sum_{\langle i,j \rangle} S_{i}^{+}S_{j}^{-} 
+\frac{U}{2}\sum_{i}\left(S_{i}^{z}\right)^{2}
-\delta\mu \sum_{i}S_{i}^{z} \\
\!-\!\frac{Jn_{0}\xi}{2} \sum_{\langle i,j \rangle}
\bigr[
  S_{i}^{z}S_{i}^{+}S_{j}^{-}+S_{i}^{-}S_{i}^{z}S_{j}^{+} +
  S_{i}^{-}S_{j}^{z}S_{j}^{+}+S_{i}^{+}S_{j}^{-}S_{j}^{z}
\\
+ \xi (S_{i}^{z}S_{i}^{+}S_{j}^{-}S_{j}^{z}+
       S_{i}^{+}S_{i}^{z}S_{j}^{z}S_{j}^{-}) 
\bigl],
\end{multline}
where $\xi=\sqrt{(n_{0}+1)/n_{0}}-1$ is a measure of the `particle-hole
symmetry-breaking'. In the work of Altman and Auerbach,\cite{Altman02} $\xi$
was set to zero.

\section{Perturbation theory}
\label{chap:rspt}

For a perturbative treatment of the dynamic structure factor one starts from
the {\it pure} Mott state $|\tilde 0\rangle$ where all sites are occupied by
exactly $n_{0}$ particles. A consistent expansion of
(\ref{eqn:structurefactor}) in $J/U$ is obtained by an admixture of virtual
particle hole pairs in the ground state
\begin{equation}
|0\rangle^{(1)} =-J \sum_{i\neq j}
\sum_{\langle l,m \rangle} |i,j\rangle 
\frac{\langle i,j| a_{l}^{\dag} a_{m}|\tilde0\rangle}{U}
+|\tilde 0\rangle,
\end{equation}
where $|i,j\rangle$ denotes a state with $n_{0}+1$ particles at $i$ and
$n_{0}-1$ particles at $j$. The excited states in (\ref{eqn:structurefactor})
are to lowest order given by the states
\begin{equation}
|n\rangle = |{\bf k}_{p},{\bf k}_{h}\rangle = 
\frac{1}{N} \sum_{i\neq j} 
e^{i({\bf k}_{p}\cdot {\bf r}_{i}-{\bf k}_{h}\cdot{\bf r}_{j})}
|i,j\rangle,
\end{equation}
with the energies
\begin{equation}
\epsilon({\bf k}_{p},{\bf k}_{h})=U
-(n_{0}+1) \epsilon_{0}({\bf k}_{p}) 
-n_{0} \epsilon_{0}({\bf k}_{h}).
\end{equation}
Inserting these perturbative states and energies into
(\ref{eqn:structurefactor}) leads to the expression
\begin{eqnarray}
\lefteqn{S^{(2)}({\bf q},\omega)=
N\left(\frac{J}{U}\right)^2 n_0(n_0+1) \times  }\\
\nonumber
&&
\int_{\rm K} \frac{d{\bf k}}{v_0} \, 
[d \gamma({\bf q}-{\bf k})- d \gamma( {\bf k})]^2
\delta(\hbar \omega-\epsilon({\bf q}-{\bf k}, {\bf k})),
\end{eqnarray}	
where $\gamma({\bf k})=1/z \sum_{l=1}^{d}\exp(i{\bf k}\cdot {\bf a}_{l})$.
Both sides of the $f$-sum rule can be calculated independently and they
coincide with a value given by
\[
\int_{0}^{\infty} d\omega \, \omega S^{(2)}({\bf q},\omega)=
4 \frac{J^{2}}{U} N d n_{0}(n_{0}+1) \left[1-\frac{1}{2}\gamma({\bf q})
\right],
\]
which shows again the quadratic dependence on $J$ and the vanishing with
$q^{2}$ for small ${\bf q}$.

\section{Effective Hamiltonian}
\label{chap:effective}

\subsection{Second order expansion}

Replacing all operators $b_{0,i}$ in (\ref{eqn:bhh}) expressed in the
$b$-bosons and collecting all terms second order in $b_{1,i}$, $b_{2,i}$ is
leading to the effective Hamiltonian
\begin{widetext}
\begin{equation}
H_{\rm eff}=2 J z
\sum_{ {\bf k} \in K}  
g_{\scriptscriptstyle 11,{\bf k}}^{\scriptscriptstyle -1} 
b_{1,{\bf k}}^{\dag} b_{1,{\bf k}} 
+ g_{\scriptscriptstyle 22,{\bf k}}^{\scriptscriptstyle -1} 
b_{2,{\bf k}}^{\dag} b_{2,{\bf k}}+
\left(
  \frac{f_{\scriptscriptstyle 11,{\bf k}}^{\scriptscriptstyle -1}}{2} 
  b_{1,{\bf k}}^{\dag} b_{1,-{\bf k}}^{\dag}
  + \frac{f_{\scriptscriptstyle 22,{ \bf k}}^{\scriptscriptstyle -1}}{2} 
  b_{2,{\bf k}}^{\dag} b_{2,-{\bf k}}^{\dag} 
  + g_{\scriptscriptstyle 12,{\bf k}}^{\scriptscriptstyle -1} 
  b_{1,{\bf k}}^{\dag} b_{2,{\bf k}}
  + f_{\scriptscriptstyle 12,{\bf k}}^{\scriptscriptstyle -1} 
  b_{1,{\bf k}}^{\dag} b_{2,-{\bf k}}^{\dag}
  + {\rm H.c.}
\right),
\end{equation}
where we extracted a factor $2Jz$ corresponding to the non-interacting
band-width from the definitions of the coefficients given by
\begin{eqnarray}
\nonumber
%g11
g_{\scriptscriptstyle 11,{\bf k}}^{\scriptscriptstyle -1} & = &
-\frac{\delta \mu}{2 J z} \sin (2\sigma)\cos(\vartheta)
+\frac{U}{4 J z}\cos(\vartheta)
+\frac{1}{2}
[ 1-\cos ^2(\vartheta ) ]
\{ n+\sqrt{n (n+1)} \cos (2 \sigma )+
       \frac{1}{2} [\sin (2 \sigma )+1] \}
\\
\nonumber
& &  
-\frac{\gamma_{\bf k}}{2}
\left\{
  \frac{1}{2} 
  [ \cos ^2(\vartheta)+1 ]
  \{ n+\frac{1}{2} [\sin (2 \sigma )+1] \}
  -\frac{1}{2} \sqrt{n (n+1)} 
  [ 1-\cos ^2(\vartheta ) ]
  \cos (2 \sigma)
\right\},
\\
%g22
\nonumber
g_{\scriptscriptstyle 22,{\bf k}}^{\scriptscriptstyle -1} & = &
+\frac{\delta \mu }{4 J z}[3-\cos (\vartheta )] \sin (2 \sigma )
+\frac{U}{8 Jz} [\cos (\vartheta )+1]
+\frac{1}{4}
[ 1-\cos ^2(\vartheta ))]
\{
 n+\sqrt{n (n+1)} \cos (2\sigma )+\frac{1}{2} [\sin (2 \sigma )+1]
\}
\\
\nonumber
& &
-\frac{\gamma_{\bf k}}{4}
\left\{
  n+\frac{1}{2} [1-\sin (2 \sigma )]
\right\}
[\cos (\vartheta )+1],
\\
%g12
\nonumber
g_{\scriptscriptstyle 12,{\bf k}}^{\scriptscriptstyle -1} & = &
-\frac{\delta \mu}{2 J z}\cos(\vartheta/2)\cos (2 \sigma )
+\cos (\vartheta/2)
\frac{1}{8} 
  [1-\cos (\vartheta )]
  [\cos (2 \sigma )+2\sqrt{n (n+1)} \sin (2 \sigma )]
\\
\nonumber
& &
-\frac{\gamma_{\bf k}}{8} 
\cos(\vartheta/2)
\left\{
  [\cos (\vartheta )+1] \cos (2 \sigma )+
  2\sqrt{n (n+1)} [1-\cos (\vartheta )] \sin (2 \sigma )
\right\},
\\
\nonumber
%f11
f_{\scriptscriptstyle 11,{\bf k}}^{\scriptscriptstyle -1} & = &
\frac{\gamma_{\bf k}}{4}
\left\{
  [1-\cos ^2(\vartheta )]
  \{n+\frac{1}{2} [\sin (2 \sigma )+1]\}
  -\sqrt{n (n+1)}\left[\cos ^2(\vartheta )+1\right] \cos(2 \sigma)
\right\},
\\
%f22
\nonumber
f_{\scriptscriptstyle 22,{\bf k}}^{\scriptscriptstyle -1} & = &
\frac{\gamma_{\bf k}}{4}
\sqrt{n (n+1)} [\cos (\vartheta )+1] \cos (2 \sigma ),
\\
%f12
f_{\scriptscriptstyle 12,{\bf k}}^{\scriptscriptstyle -1} & = &
\frac{\gamma_{\bf k}}{8} 
\cos(\vartheta/2)
\left\{
  [1-\cos (\vartheta )] \cos (2 \sigma )+
  \sqrt{n (n+1)} [\cos (\vartheta )+1] \sin (2 \sigma)
\right\},
\end{eqnarray}
where we defined again $\gamma({\bf k})=1/z \sum_{l=1}^{d}\exp(i{\bf k}\cdot
{\bf a}_{l})$.  \end{widetext}

\subsection{Dispersion in the superfluid phase}

In the superfluid phase the excitation energies are given by
\begin{equation}
\label{eqn:lengthy}
\epsilon_{s(m)}({\bf k})=Jz\sqrt{2\left[A_{\epsilon}({\bf k})\mp 
\sqrt{A_{\epsilon}({\bf k})^{2}-4B_{\epsilon}({\bf k})}\right]},
\end{equation}
and the coefficients are defined as
\begin{eqnarray*}
A_{\epsilon}({\bf k}) \!\!& = &\!\!
(g_{\scriptscriptstyle 11,{\bf k}}^{\scriptscriptstyle -1})^{2}+
(g_{\scriptscriptstyle 22,{\bf k}}^{\scriptscriptstyle -1})^{2}-
(f_{\scriptscriptstyle 11,{\bf k}}^{\scriptscriptstyle -1})^{2}-
(f_{\scriptscriptstyle 22,{\bf k}}^{\scriptscriptstyle -1})^{2}+
 \\
& & 2(g_{\scriptscriptstyle 12,{\bf k}}^{\scriptscriptstyle -1})^{2}-
2(f_{\scriptscriptstyle 12,{\bf k}}^{\scriptscriptstyle -1})^{2},
 \\
B_{\epsilon}({\bf k}) \!\!& = &\!\!
((g_{\scriptscriptstyle 11,{\bf k}}^{\scriptscriptstyle -1}-
f_{\scriptscriptstyle 11,{\bf k}}^{\scriptscriptstyle -1})
(g_{\scriptscriptstyle 22,{\bf k}}^{\scriptscriptstyle -1}-
f_{\scriptscriptstyle 22,{\bf k}}^{\scriptscriptstyle -1})-
(g_{\scriptscriptstyle 12,{\bf k}}^{\scriptscriptstyle -1}-
f_{\scriptscriptstyle 12,{\bf k}}^{\scriptscriptstyle -1})^{2})
\times
 \\
& &
((g_{\scriptscriptstyle 11,{\bf k}}^{\scriptscriptstyle -1}+
f_{\scriptscriptstyle 11,{\bf k}}^{\scriptscriptstyle -1})
(g_{\scriptscriptstyle 22,{\bf k}}^{\scriptscriptstyle -1}+
f_{\scriptscriptstyle 22,{\bf k}}^{\scriptscriptstyle -1})-
(g_{\scriptscriptstyle 12,{\bf k}}^{\scriptscriptstyle -1}+
f_{\scriptscriptstyle 12,{\bf k}}^{\scriptscriptstyle -1})^{2}).
\end{eqnarray*}

\subsection{Massive mode weight the hopping correlator} The weight $W_{\rm
mass }$ of the delta peak at $\Delta_{m}$ in $S^{\rm kin}$ is given by
\begin{multline}
W_{\rm mass}= \sin(\vartheta)^{2} \bigg[\cos(\vartheta/2)^{2} (N_{21}(0)+
P_{21}(0))-\\
\nonumber
\;\;
\frac{1}{2} \cos(\vartheta)[\sqrt{n_{0}}
+\sqrt{n_{0}+1}]^{2}(N_{11}(0)+P_{11}(0))\bigg]^2.
\end{multline}
%

%%%%%%%%%%%%%%%%%%%%%%%%%%%%%%%%%%%%%%%%%

\bibliographystyle{apsrev}
\bibliography{Refbib/ref,comments}

\begin{thebibliography}{47}
\bibitem[{com({\natexlab{b}})}]{comment:isotropy}
\bibinfo{note}{{For an isotropic lattice modulation, the expansion of the
  constraint can be avoided as the operator $T$ can be expressed as the
  difference between $H$ and local terms.}}
\bibitem[{com({\natexlab{c}})}]{comment:dm}
\bibinfo{note}{{The method of Dyson and Maleev,\cite{Dyson56,Maleev58} which is
  known to give better approximations for spin-waves in ferromagnets than the
  Holstein-Primakoff approach, is not applicable in our case due to the lack of
  the small parameter $1/S$, where $S$ is the magnitude of the spin.}}
\end{thebibliography}
 
\end{document}